\documentclass[aps,twocolumn,showpacs,floatfix,amssymb,preprintnumbers,pre]{revtex4-1}

\usepackage[dvips]{graphicx}
\usepackage{dcolumn}
\usepackage{bm}
\usepackage[dvips]{epsfig}
\usepackage{amssymb}
\usepackage{latexsym}
\usepackage{hyperref}
\usepackage{ulem}  
\usepackage{color}
\usepackage{array}
\usepackage{placeins}

\newcolumntype{C}{>{\centering}m{.1\columnwidth}}
\newcolumntype{D}{>{\centering}m{.15\columnwidth}}

\newcommand \be {\begin{equation}}
\newcommand \ee {\end{equation}}
\newcommand \bea {\begin{eqnarray}}
\newcommand \eea {\end{eqnarray}}

\begin{document}
\title{Faraday waves over a permeable rough substrate}

\author{Diego Barba Maggi$^{1,3}$, Alejandro Boschan$^1$, Roman Martino$^1$, Marcelo Piva$^1$ and  Jean-Christophe G\'eminard$^2$.}
\affiliation{$^1$Grupo de Medios Porosos, Fac. de Ingenier\'{\i}a, Universidad de Buenos Aires. Paseo Col\'{o}n 850, (C1063ACV) Buenos Aires, Argentina.}
\affiliation{$^2$Universit\'e de Lyon, Laboratoire de Physique, Ecole Normale Sup\'erieure de
Lyon, CNRS, UMR 5672, 46 All\'ee d'Italie, 96007 Lyon, France.}
\affiliation{$^3${Escuela Superior Polit\'ecnica de Chimborazo, ESPOCH, Panamericana Sur Km 1 1/2, Riobamba, Ecuador.}}

\begin{abstract}
We report on an experimental study of the Faraday instability in a vibrated fluid layer situated over a permeable and rough substrate,
consisting either of a flat solid plate or of woven meshes having different openings and wire diameters, open or closed (by a sealing paint).
We measure the critical acceleration and the wavelength (on the images from top)
at the onset of the instability for vibration frequencies between 28 and 42 Hz.
We observe that, in comparison with the flat plate, a mesh leads to an increase of the critical acceleration, whereas the wavelength is not significantly altered in none of the explored cases.
In order to rationalize the observations, we use the linear theory written for the case of a flat bottom and a viscous fluid to define an effective thickness of the fluid layer, 
which permits to define a slip length at the bottom.
For the closed meshes the slip length is simply a linear function of the distance between wires constituting the mesh,
whereas it exhibits a more complex behavior for the open meshes. 
We propose a qualitative understanding for the observed features.

\end{abstract}

\maketitle

\section{Introduction}
\label{introduction}

The instability of a fluid layer in a vertically vibrated vessel was first studied by Faraday \cite{faraday1831},
and became one of the most referenced examples of pattern formation in out-of-equilibrium, non-linear, systems \cite{cross93}.
Above a critical acceleration $a_c$, under certain conditions, a sub-harmonic instability develops in the form of standing waves at half the driving frequency \cite{douady90,bechhoefer95}, leading to a pattern of non-linear standing waves \cite{miles75}.
For inviscid fluids, the phenomenon relies in a competition between the destabilizing vibration and restoring gravity and surface tension forces \cite{benjamin54}.
In the late 90's, the theory was extended to weak viscous fluids by using a linear analysis \cite{beyer95,kumar96} and then to more dissipative systems \cite{cerda97}: the dissipation introduces a damping coefficient that depends explicitly on the layer thickness as opposed to the inviscid case.
The damping of these waves has its origin in the viscous dissipation in the free surface, bulk, bottom and side boundaries, and in the hysteresis associated with the meniscus surrounding the free surface \cite{miles67}.
The selection of the wave pattern at the free surface of the vibrated fluid
motivated numerous experimental \cite{edwards94,douady88,kudrolli96}
and theoretical studies \cite{zhang97,chen97,chen99,perinet09,perinet12},
including compound excitation \cite{edwards94,besson96}.
Specific studies of the effects of surface tension \cite{perlin00} and fluid rheology \cite{wagner99,raynal99}, of the interaction between boundary layer and bulk  \cite{perinet17},
of the streaming flows arising from dissipation \cite{vega01,martin02,vega04}, and of the effects of the side walls \cite{batson13}
are available in the literature.

The critical acceleration, $a_c$, was shown to be extremely sensitive to the thickness $h$ of the fluid layer \cite{muller98}.
For thin layers, as $h$ decreases, $a_c$ increases drastically due to the dissipation in the bottom boundary (substrate).
This strong dependence promoted analytical \cite{osipov01} and experimental studies \cite{wang06,wang07,feng16},
including the case of corrugated substrates.
It was shown that strong interaction may exist between the corrugation and the waves, up to the existence
of a forbidden band for the wavelenghts that are close to the typical horizontal scale of the corrugation.

In \cite{feng16}, the substrate consists of a square lattice of grooves
and the variation in $a_c$, compared with a flat substrate, is interpreted
in terms of the effective thickness of the fluid layer, that depends on the groove depth.
In the present work, we report on the Faraday instability in a fluid layer situated over woven meshes
having different openings and wire diameters.
The meshes present both roughness and permeability, which complements previous experimental works.


We first describe the experimental device and protocols (Sec.~\ref{sec:experimental})
and then report the experimental results (Sec.~\ref{sec:results}).
The experimental findings are then thoroughly discussed in terms of an effective thickness
of the fluid layer (Sec.~\ref{sec:analysis}) before we draw conclusions (Sec.~\ref{sec:conclusions}).

\section{Experimental setup and protocols}

The principle of the experiment is to vertically vibrate a thin layer of fluid situated over a mesh, that is either used as is (open) or coated with a sealing paint (closed).

\label{sec:experimental}
\begin{figure}[!h]
\begin{center}
\includegraphics[width=\columnwidth]{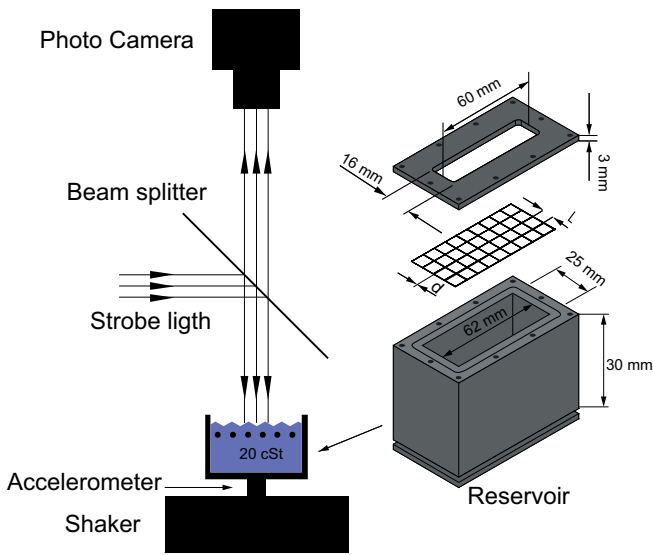}
\end{center}
\caption{(Color online)--{Sketch of the experimental setup}.
}
\label{fig:setup}
\end{figure}

The main part of the experimental setup consists of a vessel made of Poly-methyl methacrylate (PMMA), with inner dimensions 62~mm~(l) $\times$ 25~mm~(w) $\times$ 30~mm~(H)  (Fig.\ref{fig:setup}).
At top, a mesh is firmly clamped, by means of a set of iron screws, between the upper edge of the vessel and a rectangular frame (PMMA, thickness 3~mm), with an inner aperture 60~mm~(l) $\times$ 16~mm~(w).
We use woven meshes composed of interlaced iron wires, characterized by their diameter, $d$, and the distance between their centerlines, $L$.
In table~\ref{tab:grid}, we report the geometrical characteristics of the various meshes used in the experiment and, as an additional piece of information, the solidity factor $s \equiv 1-(\frac{L-d}{L})^2$ which is often used in the literature to account for the permeability \cite{lizal14}.
Depending on the experiment, the mesh is used open or closed in order to address the effects of the roughness and of the permeability; the permeability is suppressed when the mesh is closed.
In addition, we report, for comparison, data obtained for a flat solid substrate (a PMMA plate of thickness 4 mm) situated in such a way that its upper surface
would coincide with the horizontal plane intersecting all the upper mesh points.
In other words, in all cases, replacing the PMMA plate by a mesh increases the average depth of the fluid.

\begin{table}
\centering
\begin{tabular}{|c|c|c|c|}
\hline
~~~Mesh~~~&~$L$ ($mm$)~&~$d$ ($mm$)~&~~~~~$s$~~~~~  \\
\hline
\# 1 & 1.77 & 0.35 & 0.35 \\
\hline
\# 2 & 1.06 & 0.35 & 0.55 \\
\hline
\# 3 & 0.31 & 0.20 & 0.87 \\
\hline
\end{tabular}
\caption{Characteristics of the meshes used in the experiments: $L$ is the distance between wire centerlines, $d$ the wire diameter, and $s$ the solidity factor.}
\label{tab:grid}
\end{table}

The vessel is filled with silicon oil (Sigma Aldrich) with kinematic viscosity $\nu$ = 20~cSt, density $\rho$ = 0.95~g/cm$^3$, and surface tension $\gamma$ = 20.6~dyn/cm.
A digital camera (NIKON D90), with its optical axis is aligned with the vertical, takes images of the free surface (above the mesh).
A homogeneous illumination of the field of view is obtained by casting light from the source [rectangular array of white LEDs (SMD-5050) fed by a power amplifier,
Luxell Pro Line, LXP-400)] along the vertical using a beam splitter (Fig.~\ref{fig:setup}). In such configuration, the horizontal regions of the free surface
appear bright whereas the intensity of light associated to tilted regions decreases with the tilt.

The imaging system is first used to control precisely the fluid level
above the mesh. We use the fact that, as the contact line is pinned to the upper edge of the frame at top, the free surface of the fluid can be either concave or convex.
When the vessel is filled, a maximum in the average intensity of the light reflected by the free surface (and captured by the camera) corresponds to a flat and horizontal
free surface, aligned with the upper surface of the frame.
In practice, at the start of all experiments, we thus adjust the fluid volume with a micropipette (by steps of 0.2 $\mu$l), in order to get that maximum of reflected light.
From this reference situation, we remove 0.85 $\mu$l in order to avoid any overflowing during the experiments.
This procedure achieves a repeatable average thickness $h_0$ = (2.10 $\pm$ 0.02)~mm of the supernatant fluid layer over the mesh.

The vessel is attached to the vertical axis of an electromagnetic shaker (Br\"{u}el and Kjaer, V406).
The shaker is fed by a sinusoidal current from a power generator (SKP Pro Audio, MAXD-4210) driven by a function generator (GW-INSTEK, 8219A).
The resulting acceleration $a$ is measured with an accelerometer (Analog Devices, ADXL325) attached to the body of the vessel
and monitored with an oscilloscope (Gratten, GA1102CAL).
The vibration frequency $f_d$ ranges from 28 to 42~Hz for an acceleration ${{a}}$ up to 2~$g$, where $g$ is the acceleration of gravity.
The wavelength $\lambda_c$ is measured with the imaging system, having previously performed a spatial calibration.

\begin{figure}[!t]
\begin{center}
\includegraphics[width=.9\columnwidth]{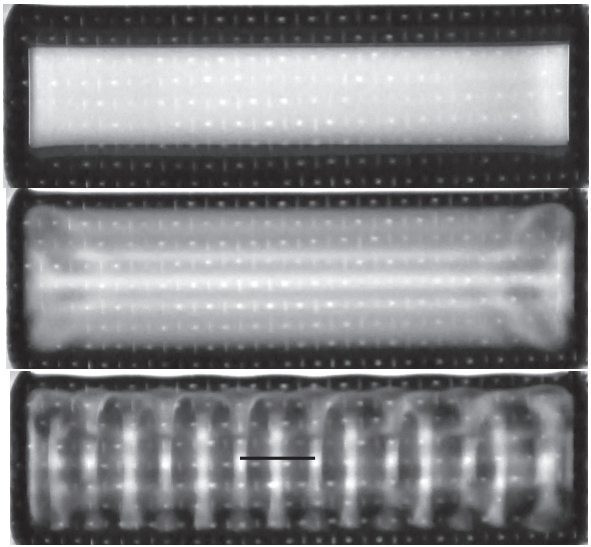}
\end{center}
\caption{Topview of the free surface of the fluid.
Top: Fluid at rest. The texture of the mesh underneath is visible (bright points).
Middle: Surface waves for $a<a_c$, below the onset of the instability (42~Hz, $a<a_c$).
Bottom: Faraday waves (42~Hz, $a = a_c$). The black bar indicates the measure of the wavelength $\lambda_c$}
\label{fig:photo}
\end{figure}

We measure the critical acceleration $a_c$ for which the subharmonic instability appears, and the corresponding wavelength $\lambda_c$ (Fig.~\ref{fig:photo}). By using a strobe light [the array of LEDs is illuminated
at half the frequency of the vertical motion (Texas Instruments, CD4040BE)],
we verified that the instability frequency matches half the value of the vibration frequency \cite{muller98}.
The experimental device and protocols were designed after the work of Douady \cite{douady90}
that sought to optimize the detection of $a_c$.
In order to check their reliability, we reproduced these experiments to an accuracy of about 2\%.
The results are reported in appendix~\ref{appendixA}.


\section{Experimental results}
\label{sec:results}

In this section we report our results in terms of the reduced critical acceleration $\Gamma_c \equiv a_c/g$, where $g$
is the acceleration of gravity, and the wavelength $\lambda_c$, characterizing the instability at its onset.

\begin{figure}[!t]
\begin{center}
\includegraphics[width=.95\columnwidth]{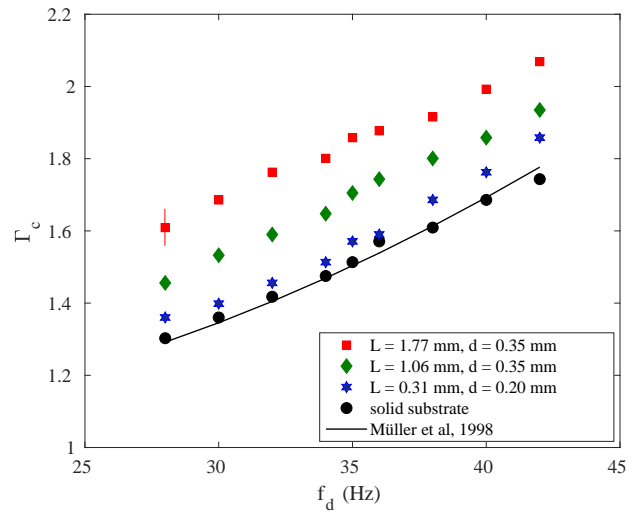}
\end{center}
\caption{Critical acceleration $\Gamma_c$ as a function of $f_d$ for the closed meshes. For the clarity of the figure, the typical error is indicated for one point only.
We observe a monotonic increase of $\Gamma_c$ with $f_d$ for any given mesh, while, at fixed $f_d$, $\Gamma_c$ increases as $L$ increases. For all meshes,  $\Gamma_c$ is greater than for the PMMA plate. The line is the theoretical predition from \cite{muller98} for a flat substrate, using  $h_0 = 2.1$~mm, $\nu = 20$~cSt, $\gamma = 20.6$~dyn/cm and $\rho = 0.95$~g/cm$^3$.
}
\label{fig:acelclosed}
\end{figure}

We first consider the case of closed meshes.
Fig.~\ref{fig:acelclosed} shows the variation of $\Gamma_c$ as a function
of vibration frequency $f_d$ for the closed meshes. Each datapoint represents
an average over three independent measurements.
On the one hand, we observe that the variation of $\Gamma_c$ with $f_d$ for the PMMA plate is in agreement with the prediction of M\"uller,
valid for a flat substrate and weak dissipation due to the fluid viscosity \cite{muller98}.
This is consistent with the estimate of the relative influence of dissipation from the dissipation parameter $\epsilon=\nu k_c^2/(\pi f_d) \sim 0.15$ in our experiments,
indicating that dissipative effects are weak, but non negligible.
On the other hand, we observe that replacing the PMMA plate by a closed mesh leads to an increase in $\Gamma_c$ for any of the considered meshes.
We remind here that the top surface of the PMMA plate is at the same level than the horizontal plane intersecting the upper mesh points
and, thus, that the average thickness of the fluid layer is increased when the PMMA plate is replaced by a mesh.
If a simple geometrical effect was at play, one would rather expect a decrease in $\Gamma_c$.
In addition, we observe that the increase in $\Gamma_c$ is larger for larger distance between the wires, $L$, even at constant wire diameter, $d$.
The effect of the mesh cannot be simply accounted by the change in the average depth of fluid [here $h_0 + d (1-s)$ where $s$ is the solidity factor].


\begin{figure}[!t]
\begin{center}
\includegraphics[width=.95\columnwidth]{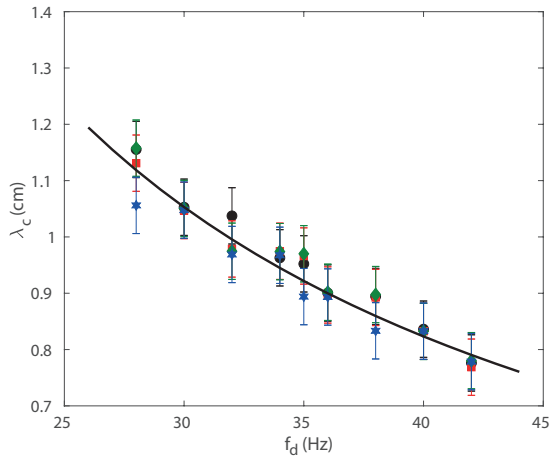}
\end{center}
\caption{Wavelength $\lambda_c$ at the instability onset as a function of $f_d$ for the closed meshes
[The symbols are the same than in Fig.\ref{fig:acelclosed} - The solid line is the dispersion relation of Eq.~(\ref{eq:lambda}) with $h_0 = 2.1$~mm, $\nu = 20$~cSt, $\gamma = 20.6$~dyn/cm and $\rho = 0.95$~g/cm$^3$].
}
\label{fig:closedlambda}
\end{figure}

To complement the experimental results, we report measurements of the wavelength $\lambda_c$ in Fig.~\ref{fig:closedlambda}.
To within the experimental accuracy (of about 1~mm), we do not observe any significant dependence of the wavelength $\lambda_c$ on the characteristics of the mesh.
Notably, the dependence of $\lambda_c$ on the frequency $f_d$ is well described by the dispersion relation of free surface waves in the inviscid case (Eq.\ref{eq:lambda},
with $h_0 = 2.1$~mm, $\nu = 20$~cSt, $\gamma = 20.6$~dyn/cm and $\rho = 0.95$~g/cm$^3$) as suggested in~\cite{muller98}. 

\begin{equation}
(\pi f_d)^2 = k_c \, \biggl( g + \frac{\gamma}{\rho} k_c^2 \biggr) \, \tanh{(k_c h_0)}
\label{eq:lambda}
\end{equation}
where $k_c \equiv 2\pi/\lambda_c$.
Taking the dissipation into account would lead to a shorter wavelength of less than 10$^{-5}$ in relative value.

\begin{figure}[!t]
\begin{center}
\includegraphics[width=.95\columnwidth]{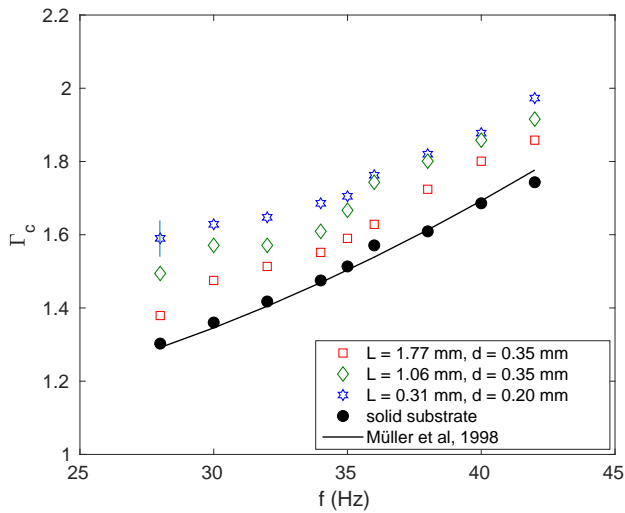}
\end{center}
\caption{Critical acceleration $\Gamma_c$ as a function of $f_d$ for the open meshes. For all meshes,  $\Gamma_c$ is greater than for the PMMA plate. The same monotonic increase of $\Gamma_c$ with $f_d$ for any mesh is observed, while for fixed $f_d$, $\Gamma_c$ increases as $L$ decreases (The error bar and the solid line are as in Fig.~\ref{fig:acelclosed}).
}
\label{fig:gamma_open_meshes}
\end{figure}

We now consider the case of the open meshes.
We again observe in Fig.~\ref{fig:gamma_open_meshes} that $\Gamma_c$ is smaller for the PMMA plate than for any mesh.
However, in contrast with what was observed for the closed meshes, the critical acceleration $\Gamma_c$ decreases as $L$ is increased, even if the wire diameter $d$ remains constant.

\begin{figure}[!t]
\begin{center}
\includegraphics[width=.95\columnwidth]{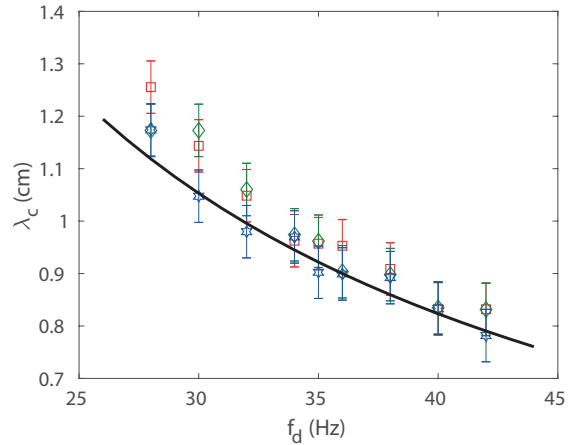}
\end{center}
\caption{Wavelength $\lambda_c$ as a function of $f_d$ for the open meshes
[The symbols are the same than in Fig.\ref{fig:gamma_open_meshes} - The solid line is as in Fig.~\ref{fig:closedlambda} .
}
\label{fig:lambda_open_meshes}
\end{figure}

Considering the wavelength, we observe that the change in the bottom properties has little effect even if slight increase of $\lambda_c$ (less than 1~mm in average) with respect to the PMMA plate is observed in Fig.~\ref{fig:lambda_open_meshes}.
However, the experimental accuracy does not make it possible to discriminate separately the effects of $L$ and $d$ on~$\lambda_c$ in this case.

\section{Analysis}
\label{sec:analysis}

To initiate the discussion, let us first summarize the experimental observations.
In short, we observe in our experimental conditions that the critical acceleration
$\Gamma_c$ for a mesh at the bottom is always larger than the one
measured for a solid substrate. The conclusion holds true the mesh being permeable (open)
or not (closed) (Figs.~\ref{fig:acelclosed}~\&~\ref{fig:gamma_open_meshes}).
The effect is significant whereas the critical wavelength $\lambda_c$ is not or weakly altered by the change in the conditions at bottom
(Figs.~\ref{fig:closedlambda}~\&~\ref{fig:lambda_open_meshes}).

\begin{figure}[!h]
\begin{center}
\includegraphics[width=.95\columnwidth]{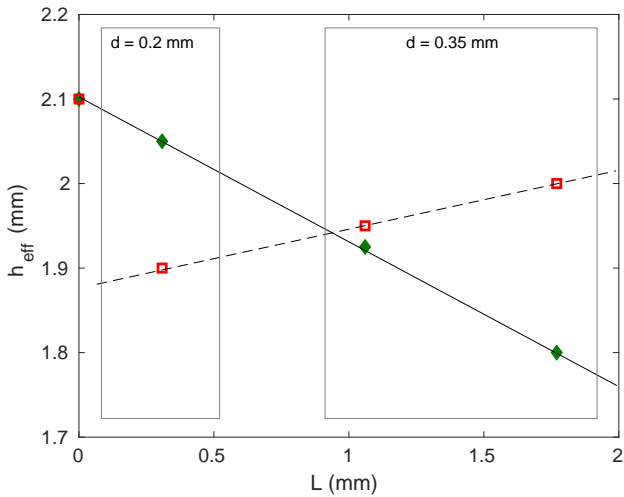}
\end{center}
\caption{Effective thickness $h_\mathrm{eff}$ vs. distance $L$ between the wires -- Full diamonds: closed mesh;
Open squares: Open mesh. The point at $L=0$ corresponds to the flat impermeable substrate.
We observe that, for the closed mesh, $h_\mathrm{eff}$ decreases linearly with $L$ ,
independently of $d$. Solid and dashed lines are linear regressions over closed and open meshes values respectively.
Note that, for the closed meshes, $h_\mathrm{eff}$ tends to $h_0$ in the limit ${L \to 0}$.}
\label{fig:effective thickness}
\end{figure}

In order to help the discussion, we propose to first rationalize our experimental results
by defining a effective thickness of the fluid layer $h_\mathrm{eff}$.
Following the idea by Feng {\it et al}~\cite{feng16}, we analyze our results 
in the light of an effective depth of liquid that participates to the instability.
However, whereas in ~\cite{feng16} the effective depth is estimated from the geometrical
characteristics of the solid bottom, we estimate $h_\mathrm{eff}$ as 
the actual depth of the fluid layer that would lead, for the same experimental
parameters and a flat bottom, to the measured critical accelation $\Gamma_c$.
To do so, we use the theoretical analysis of the Faraday instability developed by Kumar \cite{kumar96},
that is adapted to shallow layers of low viscosity fluids.
In this model, the three components of fluid velocity are assumed to vanish at the lower boundary (no slip condition).
We compute, the critical acceleration $\Gamma_c$ as function of the fluid depth $h_0$ for the values of our
experimental parameters (fluid density $\rho$, viscosity $\nu$, surface tension $\gamma$).
By interpolation of the experimental data in Figs.~\ref{fig:acelclosed}~\&~\ref{fig:gamma_open_meshes}
we get an effective fluid depth $h_\mathrm{eff}$ that is reported in Fig.~\ref{fig:effective thickness}
as function of the distance $L$ between the wires for both the open and closed meshes.
As already suggested in Sec.~\ref{sec:experimental}, $h_\mathrm{eff}$ seems to be a function of $L$, only,
as the points align in this graph for the open and the closed mesh even when $d$ is different.
We shall come back later to this viewpoint on our results.

At this point of the discussion, it is important to discuss in which regime we produce the Faraday instability
and, in particular, how far the system is from the forbidden band described by Osipov and Garc\'ia~\cite{osipov01}.
The corrugation of the substrate is expected to have strong effect when its wavelength is comparable to
the one of the developing instability , $\lambda_c$, and if the corrugation has a significant 
vertical amplitude.
In our experimental configuration, the ratio $d/h_0$ quantifies the vertical amplitude of the corrugation, yielding typically 0.1 to 0.15. With these experimental parameters, we expect the forbidden band
to be limited to a narrow region around $\lambda_c \sim L$. 
Considering now the horizontal lengthscales, we observe experimentally that the wavelength
$\lambda_c$ ranges typically from 0.8 to 1.2~cm whereas $L$ ranges from 0.31 to 1.77~mm.
The ratio $\lambda_c/L$ is thus of the order of 10.
Thus, in our experimental conditions, the Faraday instability develops in a regime far from the forbidden bands. 

In addition, it is worth estimating the typical length associated to viscosity.
In our experimental conditions, the viscous length $\sqrt{\nu/(\pi (f_d/2))}$
ranges from 0.5 to 0.7~mm in the frequency range. We remark that, while smaller
than the fluid depth $h_0$, it remains of the order of the distance $L$ between the 
wires, when meshes are used.

Replacing the solid substrate by a mesh (open or closed) drastically changes the boundary conditions 
at the bottom of the fluid. Before discussing the permeable case (open mesh), we focus on the closed
meshes. We remind here that the meshes are systematically placed below the level of the flat substrate
such that the fluid volume is increased: we estimate that the average depth of fluid below the free
surface is indeed about $h_0 + d\,(1-s)$, thus larger than $h_0$.
The question that arises here is if the corrugation of the bottom increases or decreases the volume
of liquid involved in the instability, thus the effective depth $h_\mathrm{eff}$.
The experimental results reported in Fig.~\ref{fig:effective thickness} clearly indicate that
the corrugation of the bottom leads to a decrease in $h_\mathrm{eff}$.
This conclusion is in apparent contradiction with the results obtained by Feng {\it et al} \cite{feng16} who observed that the corrugation leads to a decrease of the effective liquid mass that actually ``participates'' in the convective instability.
Note however that the authors use only one spacing $L$ (1~mm) and focus on the influence
of the corrugation amplitude in the range 54 to 270~$\mu$m for a liquid depth of about 1~mm.
Their data suggest that the appropriate effective depth would be,
using our notations, $h_0 + d/2$, but the authors note that the collapse is modest and we can remark that their data do not collapse for the largest amplitude of the corrugation. 
Our experimental data rather suggest that
$h_\mathrm{eff} = h_0 - \alpha L$ with $\alpha \simeq 0.17$ (Fig.~\ref{fig:effective thickness}). 
Both results must be compared with caution, and we conclude that they're not 
incompatible. Indeed, our experiments take place in slighly different experimental conditions
beucause the ratio $d/L$ ranges between 0.05 and 0.25 in \cite{feng16}, whereas $d/L$ ranges
between 0.15 and 0.33 in the present work. Our results are obtained in the limit
in which the effective depth proposed by Feng {\it et al} seems to work less.
There is no doubt that $h_\mathrm{eff}$ should be a function of both $d$ and $L$
but, limiting the discussion to the experimental results of the present study,
we can conclude that $h_\mathrm{eff} = h_0 - \alpha L$ in our experimental conditions
($d/L$ in the range from 0.15 to 0.33).

We can thus wonder how to explain the linear dependence of the effective fluid
depth on the horizontal scale of the mesh.
Several studies treated the slip boundary condition at the interface between a fluid and a rough solid surface composed
of a periodical arrangement of grooves. 
Hocking concluded that the nature of this condition depends much more on the distance between grooves than on their depth \cite{hocking76},
which is in agreement with the fact that $h_\mathrm{eff}$, and thus $\Gamma_c$, mainly depends on $L$.
Moreover, Niavarani {\it et al} showed that inertial effects promote the formation of recirculation zones that, in turn,
lead to an effective negative slip length. The size of the vortices scaling with the horizontal size of the grooves,
the recirculation is again consistent with the linear decrease of the effective thickness $h_\mathrm{eff}$ when $L$ is increased~\cite{niavarani09}.
This mechanism however certainly necessitates that amplitude of the corrugation be large enough, which probably explains
that the conclusion does not hold when $d/L$ is too small.

In conclusion, for the closed meshes, the corrugation leads to a decrease of the effective depth of the fluid layer (thus, of the volume of fluid that participates to the instability). This effect might be associated to enhanced dissipation
in the fluid layer at the boundary, which may contribute to the increase
of the critical acceleration $\Gamma_c$ due to the additional energy required to trigger the instability.
Several authors \cite{muller98,rajchenbach15} showed that a dissipation term related to the viscous
damping in the bottom boundary layer exists and becomes more important as $h_0$ and $f_d$ are decreased.
However our experimental device does not make it possible to observe the increase of $\Gamma_c$
at low frequencies and to isolate the contribution of dissipation, alone.

We know focus on the open meshes. 
We observe in Fig.~\ref{fig:gamma_open_meshes} that the critical accelation is
systematically larger for the mesh than for the flat bottom.
This effect could be due either to: \\
(a) a decrease of the volume of fluid that participates to the instability.\\
or \\
(b) the existence of enhanced dissipation at the bottom.\\
In contrast to what was observed for the closed meshes, the effective fluid
depth $h_\mathrm{eff}$ increases (linearly) with $L$ and $h_\mathrm{eff}$ does
not tend to $h_0$  for $L \to 0$ (Fig.~\ref{fig:effective thickness}).
Having in mind that the viscous length is of the order of 0.5 to 0.7~mm,
close to the distance $L$, the fact that $h_\mathrm{eff}$ does not tend to $h_0$  for $L \to 0$,
rules out that (a) is the only responsible for the observed trend. In the limit $L \to 0$,
the shift of the critical acceleration to larger values is necessarily due to (b).
The flow through a permeable medium (the mesh) involves energy dissipation \cite{hasimoto58}, and thus 
an increase of $\Gamma_c$, provided that the fluid can effectively flow through the mesh.
We expect this to occur in ur experiments, since the viscous length is of the order of the distance $L$.

The increase of $h_\mathrm{eff}$ with $L$ can be explained by the combination of two effects:
(a) if recirculation is at play, the size of the vortices scales like $L$ but their center
can be located deeper in the fluid layer (for a closed mesh they are necessarily located above the mesh),
at a distance from the free surface that scales like $L$; (b) the dissipation due to the flow through the grid
decreases when $L$ is increased. 

The above arguments provide a coherent picture of the mechanisms
that could explain our observations of a systematic increase
of the critical acceleration $\Gamma_c$ when a flat solid plate
is replaced by a mesh, open or closed.

\section{Conclusions}
\label{sec:conclusions}

We have studied the effects of a permeable rough substrate
on the critical acceleration $\Gamma_c$ and on the wavelength $\lambda_c$ that characterizes the onset of the Faraday instability. We observed that, in all cases, the presence of a mesh leads to larger $\Gamma_c$ compared with the flat solid plate. Regarding the influence of the mesh coarseness, taken into account through the distance between wires $L$, the acceleration $\Gamma_c$ increases with $L$ for the closed meshes (as the mesh becomes coarser), while it decreases for the open meshes. 
Considering viscous dissipation, a suitable interpretation can be found under the hypothesis that geometrical effects of recirculations prevail for the closed meshes,
while friction due to flow through the mesh dominates for the open meshes.
A complete understanding of the observations would require a proper account of the 
flow structure around (or through) the mesh and/or to introduce the adequate stress and velocity
continuity conditions at the mesh \cite{silin11} in the numerical or analytical models \cite{kumar96,muller98}.

\begin{acknowledgments}
The authors acknowledge the support from CAFCI and SENESCYT.
We thank M.~Rosen for her support, and G.~Bongiovanni for his help in the electronic settings used to measure $\Gamma_c$.
\end{acknowledgments}
~\\
\appendix
\section{Setup validation}
\label{appendixA}

In order to check the reliability of the experimental device and protocols,
we reproduced the experiments reported by Douady \cite{douady90}, by measuring the critical acceleration $a_c$ for a flat solid bottom in the same experimental conditions than the ones employed by that author, in particular, the same fluid and the same thickness $h_0 = 2.25$~mm of the fluid layer.

\begin{figure}[!b]
\begin{center}
\includegraphics[width=\columnwidth]{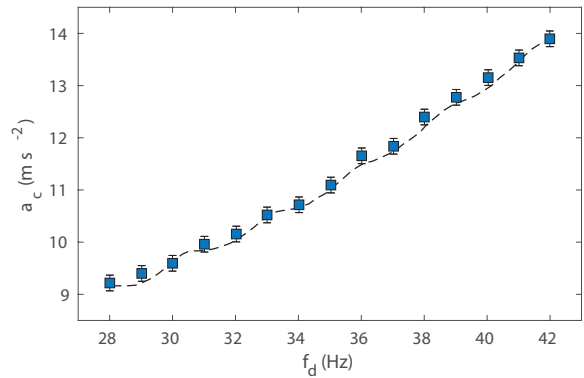}
\end{center}
\caption{Critical acceleration $\mathrm{a}_c$ as a function of the driving frequency $f_d$. Dashed line: results from Douady \cite{douady90}. Squares: this work
(error bars: $\pm 2~\%$).
[$h_0 = 2.25$~mm, $\nu = 20$~cSt, $\gamma = 20.6$~dyn/cm and $\rho = 0.95$~g/cm$^3$].
}
\label{fig:douady-acceleration}
\end{figure}

The onset of the instability was detected by using a checkpoint protocol.
To avoid hysteresis effects, the measurements were always performed for increasing
acceleration $a$ and repeatability was carefully verified.
The agreement observed in Fig.~\ref{fig:douady-acceleration} validates
our experimental device and protocols.
We estimate the error in $a_c$ to be about 2\%, at maximum.

\end{document}